\documentclass[doublecol]{epl2}

\title{Wilson ratio in Yb-substituted CeCoIn$_5$}

\author{C. Capan\inst{1} \and G. Seyfarth\inst{1,2} \and D. Hurt\inst{1} \and B. Prevost\inst{3} \and S. Roorda\inst{3} \and A. D. Bianchi\inst{3} \and Z. Fisk\inst{1}}

\institute{
  \inst{1} Department of Physics and Astronomy, University of
California Irvine, Irvine CA 92697-4575\\
  \inst{2} Department of Condensed Matter Physics, University
of Geneva, 24 quai Ernest-Ansermet, 1211 Geneva 4, Switzerland\\
  \inst{3} Department de
Physique, Universite de Montreal, Montreal H3C 3J7 Canada

 } \pacs{74.25.Dw}{Superconductivity phase diagrams}
\pacs{74.70.Tx}{Heavy fermion superconductors}
\pacs{74.62.Dh}{Effects of crystal defects, doping, substitution}

\abstract{We have investigated the effect of Yb substitution on the
Pauli limited, heavy fermion superconductor, CeCoIn$_5$. Yb acts as
a non-magnetic divalent substituent for Ce throughout the entire
doping range, equivalent to hole doping on the rare earth site. We
found that the upper critical field in (Ce,Yb)CoIn$_5$ is Pauli
limited, yet the reduced (H,T) phase diagram is insensitive to
disorder, as expected in the purely orbitally limited case. We use
the Pauli limiting field, the superconducting condensation energy
and the electronic specific heat coefficient to determine the Wilson
ratio ($R_{W}$), the ratio of the specific heat coefficient to the
Pauli susceptibility in CeCoIn$_5$. The method is applicable to any
Pauli limited superconductor in the clean limit.}

\begin{document}

\maketitle

\section{Introduction}
\indent Heavy fermion (HF) systems have been an ideal playground for
investigating unconventional superconductivity (SC) since the
discovery of SC in this class of materials\cite{PhysRevLett.43.1892,
PhysRevLett.50.1595}. Much of the attention has been focused on the
symmetry of the superconducting order parameter and the
interplay/competition between SC and
magnetism\cite{1998Natur.394...39M}. Such investigations laid ground
for magnetism as the origin of Cooper pairing, since SC in HF seems
to occur invariably in close proximity to a $T=0$ magnetic
instability\cite{2007Natur.450.1177M}. Their large specific heat
($C$) anomaly at the SC transition and the absence of SC in the
non-magnetic La analogs imply that Cooper pairs form out of heavy
quasiparticles (QP). Thus, the heavy mass and SC appear to originate
from the same mechanism. Yet, the relationship between SC and the
Kondo lattice physics, at the heart of the HF problem, has only
recently come to spotlight. In particular, a new model of
superconductivity for the 115 family of heavy fermions shows that
the composite heavy quasiparticles form only when the system becomes
superconductor\cite{Flint2008}.

\indent CeCoIn$_5$ is an ambient pressure SC with
$T_c=2.3K$\cite{2001JPCM...13L.337P} and has the unique feature of
an antiferromagnetic (AFM) quantum critical point located near the
upper critical field H$_{c2}$, indicating that AFM is superseded by
SC\cite{bianchi:257001, PhysRevLett.91.246405}. Moreover, the change
of the SC transition from second to first order for transition
temperatures $T_c \leq T^{\ast} \sim 0.7~T_{c0}$
\cite{bianchi:137002} combined with the discovery of a second SC
phase in the large B/T region of the phase diagram lead to the
suggestion that CeCoIn$_5$ is the first
realization\cite{2003Natur.425...51R, bianchi:187004} of a
Fulde-Ferrell-Larkin-Ovchinnikov (FFLO) state\cite{PhysRev.135.A550,
LarkinOvchinnikov}. Subsequent NMR measurements have shown that only
some of the NMR lines broadened within the second SC phase,
consistent with local moment ordering\cite{young:036402}. A more
recent neutron diffraction experiment has found that this second SC
phase carries AFM order that collapses at the same upper critical
field at which SC is destroyed\cite{M.Kenzelmann09192008}.
CeCoIn$_5$ is thus the first example of magnetic order being
stabilized by superconductivity, suggesting a ground state differing
from the one proposed by FF and LO.

\indent The unconventional $d_{x^2-y^2}$ gap symmetry in CeCoIn$_5$
has been established based on angular dependence of
$C$\cite{0953-8984-16-3-L02} and thermal
conductivity\cite{PhysRevLett.87.057002, vorontsov:237001}, as well
as point contact spectroscopy\cite{park:177001}. Recently, a
resonance peak has been discovered below $T_c$ in inelastic neutron
scattering\cite{stock:087001}, suggesting a magnetically mediated
pairing in analogy with the high-$T_c$ cuprates. CeCoIn$_5$ has also
been a model system for investigating the emergence of coherence in
a Kondo lattice. A phenomenological two-fluid model has been
successfully used to describe the crossover from single-ion Kondo
behavior to coherent heavy fermion ground state in Ce$_{1-x}$
La$_x$CoIn$_5$\cite{PhysRevLett.92.016401}.

\indent Here we report a specific heat investigation of CeCoIn$_5$
as a function of Yb substitution to Ce. Yb acts as a non-magnetic
divalent substituent for Ce throughout the entire doping range,
equivalent to hole doping on the rare earth site. The divergence of
the specific heat in the normal state (at $H=5T$) is moderately
suppressed with Yb doping, as a result of the Ce Kondo-lattice
dilution. The upper critical field in (Ce,Yb)CoIn$_5$ appears to be
Pauli limited, as in the pure compound, yet it exhibits a scaling
expected in the purely orbital limit. To the best of our knowledge,
this is the first report that the upper critical field is
insensitive to the amount of doping in a Pauli limited
superconductor. Furthermore, we show how the Wilson ratio ($R_{W}$),
the ratio of the specific heat coefficient to the Pauli
susceptibility, can be determined from the superconducting
condensation energy in a Pauli limited superconductor. The method
yields a Wilson ratio consistent with the expected value in pure
CeCoIn$_5$.

\section{Crystal Growth and characterization}
\indent CeCoIn$_5$ crystallizes in the tetragonal HoCoGa$_5$
structure. Single crystals of Ce$_{1-x}\emph{Ln}_{x}$CoIn$_5$
($\emph{Ln}=$Yb,Lu) were grown from excess In
flux\cite{2001JPCM...13L.337P}. The lattice parameters were
determined from Rietveld refinement of powder X-ray diffraction
patterns, using Si standard. The effective concentrations were
determined with energy dispersive X-ray spectroscopy (EDS) on the
measured single crystals, as well as proton-induced X-ray emission
microprobe (PIXE) on a mosaic of crystals from the same batches. The
magnetic susceptibility was measured using a vibrating sample SQUID
magnetometer in a field of 1T or higher applied perpendicular to
[001]. The heat capacity was measured using the standard adiabatic
heat pulse technique in a $^3$He-PPMS. The resistivity ($\rho$) was
measured using the standard four wire technique with an AC
resistance bridge. Low resistance contacts were obtained by
spot-welding Au wires.

\begin{figure}
\onefigure[scale=0.5]{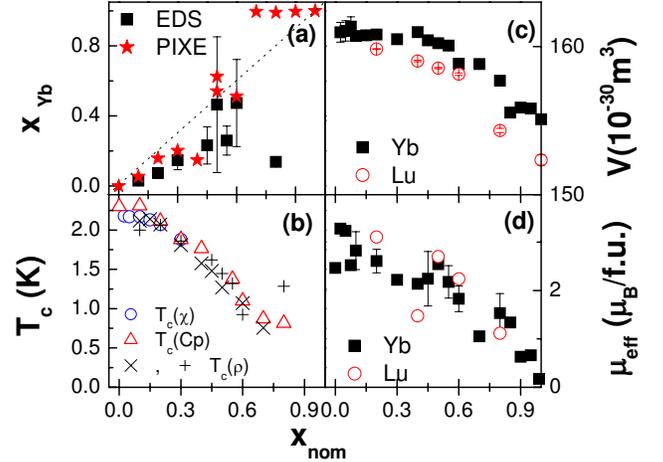} \caption{(Color online) (a)
Effective ($x_{Yb}$) vs nominal ($x_{nom}$) concentrations of Yb in
Ce$_{1-x}$ $Yb_x$CoIn$_5$, as determined from EDS and PIXE. (b)
Critical temperature $T_c$ vs $x_{nom}$ in Ce$_{1-x}$
$Yb_x$CoIn$_5$, as determined from magnetic susceptibility ($\chi$),
specific heat ($C$) and resistivity ($\rho$). (c) Lattice volume vs
$x_{nom}$ in Ce$_{1-x}$ $Ln_x$CoIn$_5$ ($Ln=$Yb, Lu), as determined
from powder X-ray diffraction. (d) Effective Curie-Weiss moment
$\mu_{eff}$ (in units of Bohr magneton) vs $x_{nom}$ in Ce$_{1-x}$
$Ln_x$CoIn$_5$ ($Ln=$Yb, Lu), as determined from $\chi$.}
\label{LattParam}
\end{figure}

\indent Fig.~\ref{LattParam} shows the doping evolution of
characteristic parameters in Ce$_{1-x}\emph{Ln}_{x}$CoIn$_5$
($\emph{Ln}=$Yb,Lu). The effective Yb concentrations, as determined
with either EDS or PIXE, are close to the nominal values for small
$x_{nom}$ but show large distribution around $x_{nom}=0.5$, as
indicated by the error bars in Fig.~\ref{LattParam}a. Phase
separation between pure YbCoIn$_5$ and Ce$_{1-x}$Yb$_{x}$CoIn$_5$ is
likely the reason why we could not reach effectively $x_{Yb} \geq
0.3$. In fact, for $x_{nom} \geq 0.7$, the batches yield essentially
pure YbCoIn$_5$, with very few Yb-substituted CeCoIn$_5$ crystals.
The difference between EDS and PIXE values reflects this difference
between a single crystal and the average concentration of the mosaic
of crystals. For simplicity, nominal concentrations will be used in
the rest of the paper. The lattice volume decreases systematically
with Yb and Lu doping due to the lanthanide contraction (see
Fig.~\ref{LattParam}c).  The Curie-Weiss moment $\mu_{eff}$ (per
formula unit), obtained from linear fits to inverse magnetic
susceptibility is suppressed below the Ce$^{3+}$ moment
($2.54\mu_B$) with Yb as with Lu doping (see Fig.~\ref{LattParam}d).
This indicates that Yb substitutes as a non-magnetic Yb$^{2+}$ ion,
resulting in a dilution of the Ce lattice. As such, the Yb doping is
\emph{not} an isoelectronic substitution, unlike La or Lu, but is
equivalent to hole doping. The absence of Curie-Weiss behavior in
pure YbCoIn$_5$ and its small Sommerfeld coefficient ($\simeq 11
mJ/K^2mol$) shows that it is not a heavy fermion.

\section{Results and Discussion}
\begin{figure}
\onefigure[scale=0.5]{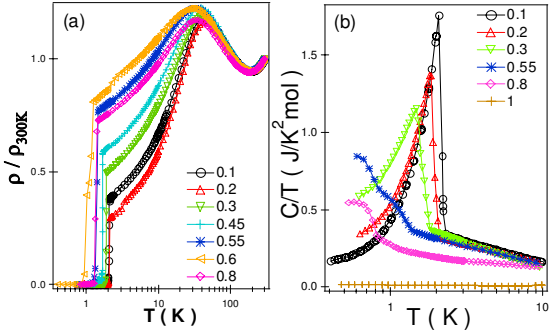} \caption{(Color online) (a)
$\rho$ vs $T$ in single crystals of Ce$_{1-x}$Yb$_x$CoIn$_5$. (b)
Electronic specific heat $\frac{C}{T}$ vs $T$ in the same crystals
at $H=0$. The indicated concentrations are nominal.}
\label{rhoandhc}
\end{figure}

\indent Figures ~\ref{rhoandhc}a and \ref{rhoandhc}b show the zero
field superconducting transition in $\rho$ and $C$ for various Yb
(EDS) concentrations. The transitions in $\rho$ are sharp, except
for $x=0.6$ which shows a double jump, indicative of an
inhomogeneous sample, consistent with the large error bars found in
the EDS (see Fig.~\ref{LattParam}a). Two different crystals have
been measured in $\rho$ from most batches and they exhibit very
similar $T_c$'s, determined from the onset of non-zero resistance,
and shown in Fig.~\ref{LattParam}b. A fairly sharp, $\lambda-$like
anomaly is observed in $C$, measured on the same single crystals for
$x=0.1, 0.2$ and $0.3$, also shown in Fig.~\ref{rhoandhc}b. The SC
anomaly is smaller and broader for $x=0.55$ and $0.8$. The onset of
the jump in $\frac{C}{T}$ defines $T_c$ for all samples. Two samples
have been measured for $x=0.55$ and both exhibit a broad jump (but
with similar $T_c$'s) possibly due to the doping inhomogeneity.
Overall, the resistive $T_c$ is in good agreement with the
thermodynamic $T_c$ determined from $\frac{C}{T}$ and $\chi$, as
shown in Figure~\ref{LattParam}b, except for $x=0.8$ with
$T_c^{\rho}>T_c^{C}$. The ratio $\frac{\Delta C}{\gamma_0 T_c}$ has
been determined with $\gamma_0$ taken as the $\frac{C}{T}$ value
linearly extrapolated to $T=0$ in the normal state at $H=5T$. In the
standard BCS theory, this ratio is expected to be 1.43 in the weak
coupling regime. For Yb concentrations $x=0.1, 0.2$ and $0.3$, the
SC appears to be bulk, in contrast to $x=0.55$ and $0.8$ with
significantly smaller $\frac{\Delta C}{\gamma_0 T_c}$ values.

\begin{figure}
\onefigure[scale=0.5]{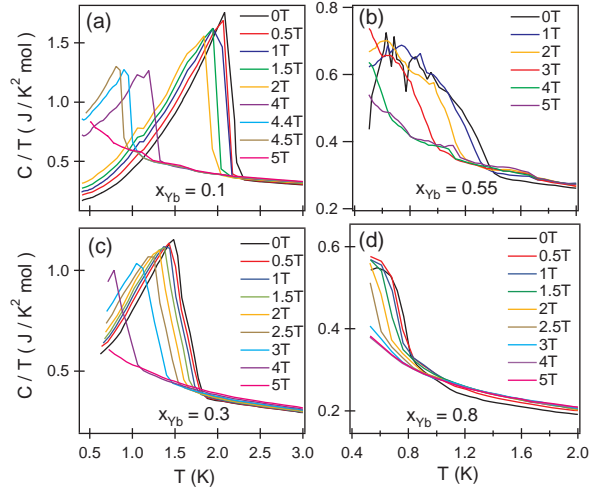} \caption{ (Color online)
Electronic specific heat $\frac{C}{T}$ vs \textbf{temperature for nominal} Yb concentrations of (a) $x_{Yb}=0.1$, (b)
$x_{Yb}=0.3$, (c) $x_{Yb}=0.55$ and (d) $x_{Yb}=0.8$, and with
magnetic fields ranging from 0 to 5~T, applied parallel to [001].}
\label{CvsT_All}
\end{figure}

\indent The temperature and field dependence of the electronic specific heat
$\frac{C}{T}$ in Yb-doped CeCoIn$_5$ is shown in
Fig.~\ref{CvsT_All}, for $x_{Yb}=0.1, 0.3, 0.55$ and $0.8$
(nominal). The electronic contribution is obtained by subtracting
the phonon contribution, determined from $C$ of YbCoIn$_5$.
$\frac{C}{T}$ has a divergent $T-$dependence down to $T_{c2}(H)$ for
all Yb concentrations, with little change in slope with increasing
magnetic field. Moreover, the divergence of $\frac{C}{T}$ extends
down to lowest $T \simeq 0.5K$ at $H=5T$ in these samples, a
characteristic shared with the pure CeCoIn$_5$. This divergence has
been attributed to a field tuned QCP near $H_{c2}^0$ in
CeCoIn$_5$\cite{bianchi:257001, PhysRevLett.91.246405}. As seen in
Fig.~\ref{rhoandhc}b and Fig.~\ref{CvsT_All}, the divergence of
$\frac{C}{T}$ becomes weaker with increasing Yb concentration and
the corresponding $\gamma_0$, obtained from linear extrapolation of
$\frac{C}{T}$ vs T at $H=5$T, decreases systematically with
$x_{Yb}$, consistent with a dilution of the Kondo lattice (see
table~\ref{tab.1}).

\begin{figure}
\onefigure{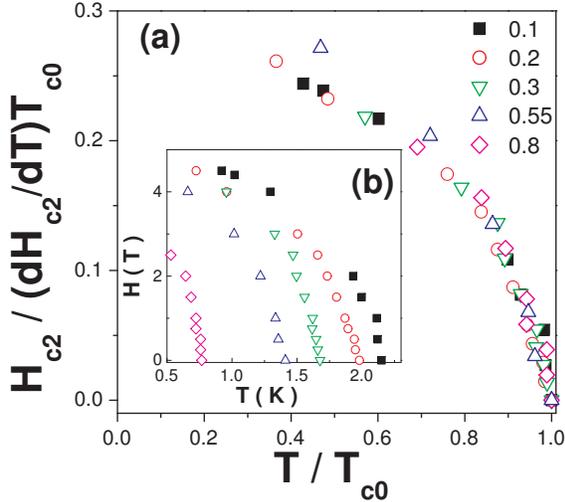} \caption{(Color online) H-T phase
diagram in Ce$_{1-x}$Yb$_x$CoIn$_5$. (a) Reduced critical field
$\frac{H_{c2}}{\frac{dH}{dT}T_c}$ vs reduced critical temperature
$\frac{T}{T_{c0}}$ for various Yb concentrations (nominal). (b)
Upper critical field $H_{c2}$ vs Temperature.} \label{Hc2}
\end{figure}

\indent Figure~\ref{CvsT_All} also shows the magnetic field
suppression of the superconducting transition, for magnetic field
applied parallel to the [001] axis, with a broader and smaller jump
as the field is increased, due to the presence of vortices. Unlike
in the pure case, the SC transition in the Yb-doped compounds does
\emph{not} sharpen into a first order anomaly in the T=0 limit, as
seen in Fig~\ref{CvsT_All}, likely due to disorder introduced by
doping\cite{PhysRev.148.362}. The corresponding H-T phase diagrams
(deduced from $C$) are shown in Fig.~\ref{Hc2}a and ~\ref{Hc2}b. The
upper critical field ($H_{c2}(T)$) is determined from the kink in
the entropy $S$, corresponding to the midpoint of the specific heat
jump seen in Fig.~\ref{CvsT_All}. $S$ is obtained by integrating
$\frac{C}{T}$, following a quadratic\cite{PhysRevLett.86.5152}
extrapolation of $C$ vs $T$ to $T=0$.  For $x_{Yb}=0.55$, the
average critical field of two samples is shown. In pure CeCoIn$_5$
the first order SC transition\cite{bianchi:137002} as well as the
$H_{c2}(T)$ fits\cite{Won:180504} give clear evidence for a Pauli
limited $H_{c2}$. Despite the absence of a first order transition in
Yb doped compounds, the Pauli limit\cite{PhysRevLett.9.266} still
applies: the extrapolation based on the standard
expression\cite{PhysRev.147.288}, $H_{c2}^0 \simeq
-0.7\frac{dH_{c2}}{dT}T_c$, yields an orbital critical field far in
excess of the observed transition field, see table~\ref{tab.1}.

\indent Figure~\ref{Hc2}a shows that the $H_{c2}$ data for all Yb
concentrations collapse onto a single curve when scaled by the
initial slope at $T_c$. Such a scaling is expected in the purely
orbital limit, as the reduced critical field vs the reduced
temperature $\frac{T}{T_c}$ is only weakly dependent on the disorder
level in this limit\cite{PhysRev.147.288}. In the Pauli limit there
is no reason to expect a similar scaling, however the orbital
mechanism is still predominant in the zero field limit.  The
implication of this scaling is that the Maki
parameter\cite{PhysRev.148.362}, the ratio of the orbital critical
field to the Pauli limiting field : $\alpha=\frac{\surd 2
H_{c2}^{0}}{H_{P}}$, is independent of $x_{Yb}$.  This is not a
trivial result, knowing that $\alpha$ decreases under
pressure\cite{miclea:117001}. $\alpha$ was estimated to be 3.6 for
$H\| [001]$ in pure CeCoIn$_5$\cite{bianchi:137002}. We have used
this value to determine $H_{P}$ in the Yb-doped samples from
$H_{c2}^0$, see table~\ref{tab.1}).

\begin{table}
\caption{Doping dependence of characteristic parameters in
Ce$_{1-x}$Yb$_x$CoIn$_5$. $x_{Yb}$ : nominal Yb concentration,
$T_c$: critical temperature at $H=0$ from heat capacity, $\gamma_0$:
the $T=0$ linear extrapolation of $\frac{C}{T}$ at $H=5T$ in
J/K$^2$mol, $U_{c}$: the superconducting condensation energy in
J/mol, $H_{c2}^{0}$(T) and $H_{P}$(T): the orbital upper critical
field and Pauli limiting field.} \label{tab.1}
\begin{center}
\begin{tabular}{cccccc}
 $x_{Yb}$ & $T_c$(K) & $\gamma_0$ & $U_{c}$ & $H_{c2}^{0}$(T) & $H_{P}$(T)\\
 0&2.3&1.2&1.43&13.6&5.3\\
 0.1&2.19&1.15&1.05&13.2&5.2\\
 0.2&1.98&0.96&0.69&12.1&4.7\\
 0.3&1.68&0.96&0.47&12.3&4.8\\
 0.55&1.45&0.84&-&10.2&4.0\\
 0.8&0.77&0.53&-&8.9&3.5
\end{tabular}
\end{center}
\end{table}

\indent In a d-wave BCS superconductor, the superconducting
condensation energy, $U_c$, is related to the specific heat jump
$\Delta C$ at the SC transition via the standard
relation\cite{CooperLoramTallon}: $U_{c}=\frac{(\Delta C) T_c}{4}$.
Since $\frac{C}{T}$ in the normal state is divergent in CeCoIn$_5$
and since it is likely a strong coupling
superconductor\cite{2001JPCM...13L.337P}, the use of this formula is
questionable. Alternatively, we have determined $U_c$ directly from
integration of $\int_{0}^{T_c} (S_n-S_s)dT$ up to $T_c$, following
extrapolation of the $\frac{C}{T}$ data to T=0.  We have excluded
the x=0.55 and 0.8 samples from the analysis since the lowest
temperature (0.4K) in the data does not allow a proper extrapolation
to T=0. The results are listed in table~\ref{tab.1}. The obtained
value of $U_c$ for pure CeCoIn$_5$ (1.43~J~mol$^{-1}$) compares
favorably with the value (1.34~J~mol$^{-1}$) determined from
magnetization\cite{Tayama}. For the doped samples, we found that
the condensation energy decreases with increasing Yb concentration,
a trend similar to the one reported for La (not shown) or Sn
doping\cite{bauer:245109}.

\indent The combination of $U_c$ and $H_{P}$ then allows the
determination of the Pauli paramagnetic susceptibility ($\chi_0$).
In fact, $H_{P}$ is related to $U_{c}$ via\cite{PhysRevLett.9.266}:
$U_{c}= \frac{1}{2}\chi_0 H_{P}^2$. This expression is
originally derived for an s-wave
superconductor\cite{PhysRevLett.9.266} and overlooks the possibility
of a finite Pauli susceptibility at T=0 in the superconducting
state\cite{AG1}. For a d-wave superconductor, it only remains valid
in the clean limit. In the presence of nodal quasiparticles in a
d-wave superconductor, there should be a contribution to the free
energy in the superconducting state and thus the Pauli critical
field should be derived from: $F_n-\chi_n H_{P}^2=F_s-\chi_s
H_{P}^2$, where the subscripts $"n"$ and $"s"$ refer to the normal
and the superconducting states. This is equivalent to $\Delta F=
F_n-F_s=(1-\frac{\chi_s}{\chi_n})\chi_n H_P^2$ and thus $U_{c}=
\frac{1}{2}(1-\frac{\chi_s}{\chi_n})\chi_n H_{P}^2$. It is
straightforward to assume that the fraction of excited nodal
quasiparticles, $\frac{\chi_s}{\chi_n}$, should be proportional to
the Yb concentration, but since we do not have an exact
determination of this ratio, we will not pursue the analysis in the
Yb doped compounds. Instead, we focus on pure CeCoIn$_5$, which is
in the clean limit, implying that $\frac{\chi_s}{\chi_n} \ll 1$. For
CeCoIn$_5$, $\chi_0=10^{-4}/4\pi$ determined from $U_c$ via this
formula is close to the value of the c-axis
susceptibility\cite{2001JPCM...13L.337P}, $\chi=1.54~10^{-4}/4\pi$
at $1.8~K$. The resulting Wilson ratio is $R_W =
\frac{\chi_0}{\gamma_0} = 0.76~R_W^0$, where $R_W^0$ is the free
electron value. In this method, the error on $R_W$ essentially comes
from the uncertainty on $U_c$ (via $\chi_0$).

The Wilson ratio of a free electron gas is defined as the ratio of
the Pauli paramagnetic susceptibility to the electronic specific
heat (Sommerfeld) coefficient and is a universal number:
$R_W^0=\frac{\chi}{\gamma}=\frac{3 \mu_B^2}{\pi^2 k_B^2}$. Despite
the large mass renormalization, heavy fermion systems exhibit Wilson
ratios very close to the free electron gas value. This is due to
the fact that the mass renormalization corresponds to an enhanced
density of states at the Fermi level, and the latter determines both
the paramagnetic susceptibility and the specific heat
coefficient\cite{Zach:Review}. The difficulty in estimating the
Wilson ratio in heavy fermion systems is associated with the
experimental determination of the low temperature Pauli
susceptibility. In fact, the magnetic susceptibility is
overwhelmingly dominated by the Curie-Weiss contribution of Ce
moments. Here we show that this difficulty can be overcome in the
case of Pauli limited heavy fermion superconductors, making use of
the superconducting condensation energy. The $R_W$ we obtained from
this method in pure CeCoIn$_5$ is close to but somewhat lower than
the expected value of $R_W=\frac{2J+1}{2J}R_W^0=1.2R_W^0$ for a
system of J=5/2 local moments (corresponding to Ce), confirming the
validity of the method.

\section{Conclusion}
\indent In conclusion, we have investigated the effect of Yb
substitution on the superconducting and heavy fermion properties of
CeCoIn$_5$. Our findings can be summarized as follows: (i) the
suppression of the condensation energy and Sommerfeld coefficient
$\gamma_0$ with Yb doping is due to the dilution of the dense Ce
Kondo lattice, (ii) the upper critical field exhibits a scaling that
is unusual for Pauli-limited superconductors, implying a doping
independent Maki parameter. Moreover, we introduce a new method for
the determination of the Wilson ratio from the superconducting
condensation energy, which is valid for any clean superconductor in
the Pauli limit. The value we estimate is consistent with the
expected value for pure CeCoIn$_5$.

\acknowledgments We acknowledge stimulating discussions with S.
Nakatsuji, J.~D. Thompson, P. Coleman, I. Vekhter and R. Movshovich.
Z. F. acknowledges Grant No. NSF-DMR-0503361. A.~D.~B. received
support from the Natural Sciences and Engineering Research Council
of Canada (Canada), Fonds Quebecois de la Recherche sur la Nature et
les Technologies (Quebec), and the Canada Research Chair Foundation.

\end{document}